\documentclass[10pt,aps,prl,twocolumn,superscriptaddress]{revtex4-2}
\usepackage{amsmath,amssymb,amsfonts,bm}
\usepackage[breaklinks=true,colorlinks,citecolor=blue,linkcolor=blue,urlcolor=blue]{hyperref}
\usepackage{graphicx}
\usepackage{txfonts}
\usepackage{braket}
\usepackage{comment}
\usepackage{siunitx}
\usepackage{algorithm}
\usepackage{algpseudocode}
\usepackage{enumitem}
\usepackage{mathtools}
\usepackage[normalem]{ulem}
\usepackage{tabularx, booktabs, multirow}
\usepackage{graphicx}
\usepackage{tikz}
\usepackage{orcidlink}

\newcommand{\trace}[1]{\text{tr}#1}

\newcommand{\eqlabel}[1]{Eq.~\eqref{#1}}
\newcommand{\figlabel}[1]{Fig.~\ref{#1}}

\newcommand{\tablabel}[1]{Table~\ref{#1}}

\begin{document}
\title{Sequential Quantum Computing}

\author{Sebastián V. Romero$^{\orcidlink{0000-0002-4675-4452}}$}
\affiliation{Kipu Quantum GmbH, Greifswalderstrasse 212, 10405 Berlin, Germany}
\affiliation{Department of Physical Chemistry, University of the Basque Country UPV/EHU, Apartado 644, 48080 Bilbao, Spain}
\author{Alejandro Gomez Cadavid$^{\orcidlink{0000-0003-3271-4684}}$}
\affiliation{Kipu Quantum GmbH, Greifswalderstrasse 212, 10405 Berlin, Germany}
\affiliation{Department of Physical Chemistry, University of the Basque Country UPV/EHU, Apartado 644, 48080 Bilbao, Spain}

\author{Enrique Solano$^{\orcidlink{0000-0002-8602-1181}}$}
\affiliation{Kipu Quantum GmbH, Greifswalderstrasse 212, 10405 Berlin, Germany}

\author{Narendra N. Hegade$^{\orcidlink{0000-0002-9673-2833}}$}
\email{narendrahegade5@gmail.com}
\affiliation{Kipu Quantum GmbH, Greifswalderstrasse 212, 10405 Berlin, Germany}
\date{\today}

\begin{abstract}
We propose and experimentally demonstrate sequential quantum computing (SQC), a paradigm that utilizes multiple homogeneous or heterogeneous quantum processors in hybrid classical-quantum workflows. In this manner, we are able to overcome the limitations of each type of quantum computer by combining their complementary strengths. Current quantum devices, including analog quantum annealers and digital quantum processors, offer distinct advantages, yet face significant practical constraints when individually used. SQC addresses this by efficient inter-processor transfer of information through bias fields. Consequently, measurement outcomes from one quantum processor are encoded in the initial-state preparation of the subsequent quantum computer. We experimentally validate SQC by solving a combinatorial optimization problem with interactions up to three-body terms. A D-Wave quantum annealer utilizing 678 qubits approximately solves the problem, and an IBM’s 156-qubit digital quantum processor subsequently refines the obtained solutions. This is possible via the digital introduction of non-stoquastic counterdiabatic terms unavailable to the analog quantum annealer. The experiment shows a substantial reduction in computational resources and improvement in the quality of the solution compared to the standalone operations of the individual quantum processors. These results highlight SQC as a powerful and versatile approach for addressing complex combinatorial optimization problems, with potential applications in quantum simulation of many-body systems, quantum chemistry, among others.
\end{abstract}

\maketitle

Composition of homogeneous and heterogeneous computing devices is central to modern classical computational architectures. This allows the integration of specialized processors such as CPUs for complex control flows, GPUs for dense linear algebra, ASICs for task-specific optimizations, FPGAs for customizable hardware acceleration, and neuromorphic chips optimized for energy-efficient pattern recognition~\cite{nickolls2008scalable,brodtkorb2010state,tomov2010towards,schuman2022opportunities,farsa2025reconfigurable}. Combining these diverse computational units consistently yields improvements in performance and energy efficiency. Quantum computing similarly encompasses diverse architectures, including analog quantum simulators, quantum annealers, digital quantum computers, and neuromorphic quantum devices~\cite{markovic2020quantum,cheng2023noisy}. These platforms employ various hardware technologies such as superconducting circuits, trapped-ion systems, photonic processors, neutral-atom arrays, and spin-based qubits. Each technology offers some drawbacks but also unique advantages, such as rapid gate operations in superconducting circuits and spin qubits, long-range connectivity in trapped-ion and neutral-atom systems, and efficiency in optimization tasks for quantum annealers~\cite{cheng2023noisy}. Integrating different quantum platforms in sequential workflows promises significant benefits via the mitigation of their individual limitations, yet this approach poses unique challenges due to the fragile nature of quantum information transfer.

Recent efforts toward quantum computing integrations include modular quantum computing, connecting identical modules via photonic or microwave links~\cite{aghaee2025scaling}, and distributed quantum computing, employing 
entanglement to interconnect distinct quantum processors~\cite{caleffi2024distributed,main2025distributed}. Techniques like parallel quantum computing~\cite{das2019case} and mid-circuit measurements~\cite{decross2023qubit} distribute algorithms across simultaneously operating quantum chips. However, these methods do not explicitly exploit the complementary advantages of homogeneous or heterogeneous quantum processors at different computational stages.

In this Letter, we introduce Sequential Quantum Computing (SQC), a framework designed to integrate homogeneous or heterogeneous quantum processors in sequential workflows (see~\figlabel{fig:sqc_schematics}). SQC transfers intermediate information of the quantum states efficiently between processors using bias fields~\cite{grass2019quantum,grass2022quantum,cadavid2024biasfielddigitizedcounterdiabaticquantum,romero2024biasfielddigitizedcounterdiabaticquantum, chandarana2025runtime, iskay}, encoding measurement outcomes from one processor as local longitudinal fields during initial state preparation of subsequent processors. We demonstrate SQC on a classical optimization problem and discuss its broader applicability, including scenarios such as quantum chemistry and materials science, where classical shadow tomography \cite{aaronson2020shadow} can augment conventional computational basis measurements.%
\begin{figure}[!t]
    \centering
    \includegraphics[width=\linewidth]{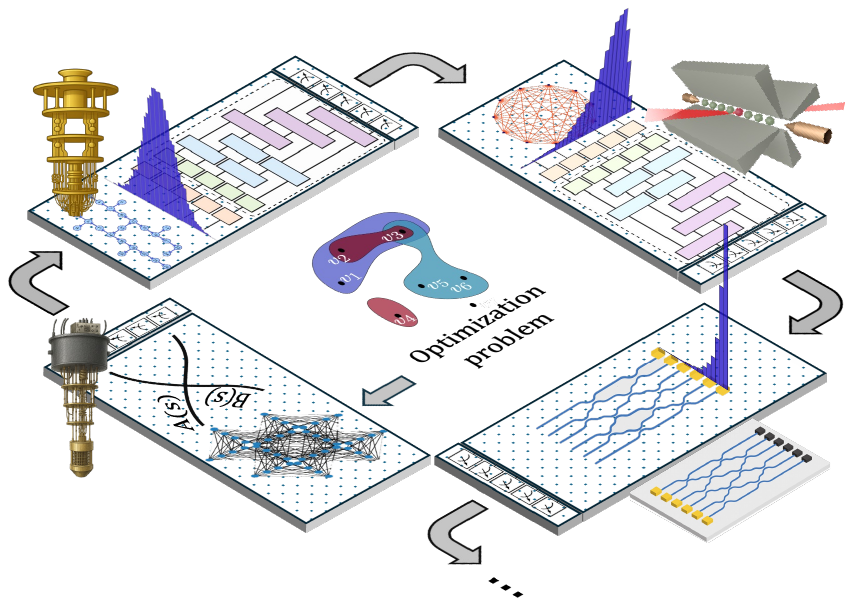}%
    \caption{Sequential quantum computing schematic. Different quantum hardware technologies offer distinct advantages. SQC integrates homogeneous or heterogeneous quantum processors, combining their strengths and addressing their limitations to improve solution quality while reducing resource usage.}\vspace{-6mm}
    \label{fig:sqc_schematics}
\end{figure}%

\emph{Formulation.}---To illustrate the potential of SQC, we consider the integration of two distinct quantum processors: an analog quantum annealer based on superconducting flux qubits and a digital quantum processor based on superconducting transmon qubits.
The first is a D-Wave Advantage~\cite{dwave} superconducting quantum annealer specialized for quadratic unconstrained binary optimization problems, described by the Hamiltonian $
H(t) = -\frac{A(t)}{2}\sum_i \sigma_i^x + \frac{B(t)}{2}\left(\sum_{i} h_i \sigma_i^z + \sum_{i<j} J_{ij} \sigma_i^z \sigma_j^z\right)$. Here, $\sigma_i^x$ and $\sigma_i^z$ are Pauli operators acting on qubit $i$, $h_i$ denote local fields, $J_{ij}$ represent spin-glass couplings, and $A(t)$ and $B(t)$ are time-dependent control parameters. Their current quantum annealer contains over 4000+ qubits with fixed connectivity and coherence times limited to tens of nanoseconds. Despite demonstrating quantum supremacy for certain quantum simulation tasks~\cite{king2025beyond}, these annealers remain restricted to stoquastic Hamiltonians, i.e., Hamiltonians whose off-diagonal elements in the standard basis are real and non-positive, which quantum Monte Carlo methods can efficiently simulate~\cite{bravyi2015montecarlo}. Nevertheless, quantum annealers have shown strong heuristic performance in combinatorial optimization~\cite{munoz2025scaling}. The second processor is IBM's superconducting digital quantum computer~\cite{ibm}, comprising 156 qubits arranged in a heavy-hexagonal lattice~\cite{chamberland2020topological}. Although universally programmable, practical applications are limited by sparse connectivity, short coherence times, and modest qubit numbers. Hence, digital quantum processors currently excel primarily with sparse computational problems.

In the following sections, we describe an SQC protocol that effectively combines the advantages of the abovementioned quantum processors. Initially, the D-Wave quantum annealer rapidly generates candidate solutions to our optimization problem. A bias-field strategy then transfers these measurement outcomes by encoding them into the input states of the IBM quantum processor. The IBM processor subsequently employs the Bias-Field Digitized Counterdiabatic Quantum Optimization (BF-DCQO) protocol~\cite{cadavid2024biasfielddigitizedcounterdiabaticquantum,romero2024biasfielddigitizedcounterdiabaticquantum, chandarana2025runtime, iskay}, incorporating approximate non-stoquastic counterdiabatic terms to further refine these solutions. Finally, we experimentally benchmark our SQC approach against solving the same problem independently on each processor.

As a concluding remark, SQC can be extended beyond classical optimization to quantum simulation and quantum chemistry by transferring intermediate quantum states or observables measured from one quantum processor to another, potentially enhancing computational efficiency and accuracy. We leave the exploration of these applications for future work. Additionally, SQC is not limited to analog-to-digital transfers. Quantum processors of various architectures, whether analog, or digital, can be cleverly combined depending on the computational problem, available hardware, and user-defined goals.

\emph{Methodology.}---To demonstrate SQC, we use a higher-order Ising  model, representative of many industrially relevant combinatorial optimization tasks, whose corresponding Hamiltonian reads as
\begin{equation}\label{eq:hubo}
H_f =\sum_i h_i^z \sigma^z_{i} + \sum_{i<j} J_{ij} \sigma^z_{i} \sigma^z_{j}+ \sum_{i<j<k} K_{ijk} \sigma^z_{i} \sigma^z_{j} \sigma^z_{k} + \cdots,
\end{equation}
where the ground state encodes the exact solution to the combinatorial optimization problem. A common approach to this problem uses adiabatic quantum optimization, which evolves a quantum system from an easily-prepared ground state of an initial Hamiltonian $H_i$ towards the target Hamiltonian $H_f$. This evolution is generated by a time-dependent Hamiltonian $H_\text{ad}(\bm{\lambda})$, with $\bm{\lambda}(t) = (\lambda_1(t), \cdots, \lambda_M(t))$ a set of $M$ time-dependent smooth scheduling functions that enforce the boundary condition $H_\text{ad}(\bm{\lambda}(0))=H_i$ at initial time and $H_\text{ad}(\bm{\lambda}(T))=H_f$ at final time $T$. In practical scenarios, the number of scheduling functions is restricted, as seen in the previous section. In the adiabatic limit, where $\dot{\bm{\lambda}}(t)\to 0$, the system's final state converges to the ground state of $H_f$. Analog quantum devices, such as quantum annealers, 
can tackle certain quadratic Ising spin-glass problems efficiently. However, solving higher-order problems requires a qubit-overhead coming from mapping the problem to a quadratic one. Alternatively, it is possible to accelerate the slow adiabatic evolution by introducing auxiliary counterdiabatic terms that suppress diabatic transitions~\cite{demirplak2003adiabatic, berry2009transitionless}. Assuming a single-schedule $\lambda(t)$, the total Hamiltonian takes the form $H_\text{cd}(\lambda)=H_\text{ad}(\lambda)+\dot{\lambda}A_\lambda$ with $A_\lambda$ the adiabatic gauge potential~\cite{kolodrubetz2017geometry}. Nevertheless, implementing the exact gauge potential is not generally possible due to its many-body structure, and it requires knowing the instantaneous spectrum. Accordingly, approximate implementations have been proposed~\cite{kolodrubetz2017geometry,sels2017minimizing,claeys2019floquet,hatomura2021controlling,takahashi2024shortcuts}, where the gauge potential is expanded as a nested-commutator series up to order $l$ as $A^{(l}_\lambda=i\sum_{k=1}^l \alpha_k(t) \mathcal{O}_{2k-1} (t)$, with $\mathcal{O}_{0}(t) = \partial_\lambda H_{\text{ad}}$ and $\mathcal{O}_{k}(t) = [ H_{\text{ad}}, \mathcal{O}_{k-1}(t) ]$. In the limit $l\to\infty$, the expansion converges to the exact gauge potential. The coefficients $\alpha_k$ are obtained by minimizing the action $S_l=\trace{[G_l^2]}$ with $G_l=\partial_\lambda H_\text{ad} - i\big[H_\text{ad},A^{(l}_\lambda\big]$. Due to non-stoquasticity~\cite{hormozi2017nonstoquastic}, its implementation on analog quantum devices is challenging. To overcome it, digitized counterdiabatic quantum optimization (DCQO)  has been proposed, which leverage the flexibility of digital quantum computers~\cite{hegade2021shortcuts}. The resulting time-evolved operator can be decomposed in $n_\text{trot}$ steps as $U(T) \approx \prod_{k=1}^{n_{\text{trot}}} \prod_{j=1}^{n_\text{terms}} \exp [-i \Delta t \gamma_j(k \Delta t) H_j]$, with $H_\text{cd}=\sum^{n_\text{terms}}_{j=1}\gamma_j(t)H_j$ decomposed in $n_\text{terms}$ different $H_j$ operators and $\Delta t=T/n_\text{trot}$. %
\begin{figure*}[!tb]
    \centering
    \includegraphics[width=\linewidth]{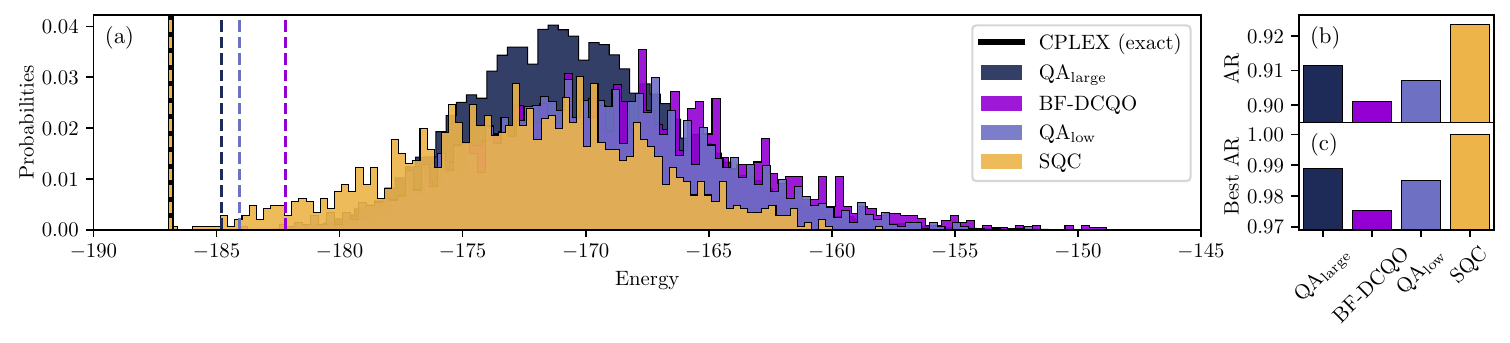}\vspace{-2mm}%
    \caption{Results for the heavy-hexagonal 156-qubit NN HUBO [\eqlabel{eq:hubo}] on quantum hardware. (a) Using D-Wave, post-processed distributions using $300000$ and $3000$ samples (dark blue and lavender, respectively) and $t_a=\SI{90}{\micro\second}$. Using IBM, best post-processed distribution after ten iterations of BF-DCQO (purple). Using both D-Wave and IBM (SQC approach, yellow), one iteration of BF-DCQO after initializing the bias fields with the post-processed D-Wave distribution of $3000$ samples. (b)-(c) Approximation ratios and best approximation ratios obtained.}\label{fig:sqc_nn_hubo}
\end{figure*}%
\begin{table*}[!tb]\vspace{-0mm}
    \caption{Performance of the instances tested [\eqlabel{eq:hubo}] under different approaches, where the best results are in bold. For IBM platforms, we assume a sampling rate of $\SI{10}{kHz}$~\cite{kotil2025quantum}. For quantum annealers, the runtime is computed as the product between the number of shots and annealing time, assuming sampling rates of few MHz.}\label{tab:performance}
    \begin{ruledtabular}\begin{tabular}{llrrrrrr}
       \multicolumn{1}{c}{Problem}  & \multicolumn{1}{c}{Approach} & \multicolumn{1}{c}{Number of shots} & \multicolumn{1}{c}{AR} & \multicolumn{1}{c}{Best AR} & \multicolumn{1}{c}{Best energy} & \multicolumn{1}{c}{Qubits} \\ \midrule
       \multirow{4}{*}{Heavy-hexagonal NN HUBO [\eqlabel{eq:hubo}]} & Standalone QA (large) & $290000$ ($\SI{26.1}{s}$) & $\bm{96.63\%}$ & $98.89\%$ & $-184.78$ & $678$ \\
        & Standalone BF-DCQO & $50000$ ($\SI{5.0}{s}$) & $90.10\%$ & $97.52\%$ & $-182.21$ & 156 \\
        & Standalone QA (low) & $3000$ ($\SI{0.6}{s}$) & $90.70\%$ & $98.51\%$ & $-184.07$ & $678$ \\
        & SQC & $\bm{8000}$ ($\SI{1.1}{s}$) & $94.95\%$ & $\bm{100\%}$ & $\bm{-186.86}$ & $678/156$ \\ 
    \end{tabular}\end{ruledtabular}\vspace{-3mm}
\end{table*}%

Building upon the DCQO protocol~\cite{hegade2022digitized} and quantum annealing (QA) with bias fields~\cite{grass2019quantum,grass2022quantum}, BF-DCQO~\cite{cadavid2024biasfielddigitizedcounterdiabaticquantum,romero2024biasfielddigitizedcounterdiabaticquantum, chandarana2025runtime, iskay} iteratively performs DCQO taking the solution from each step as an input of the subsequent iteration. In particular, we initially set $h^x_j=-1$ and $h^b_j=0$ such that $H_i=-\sum^{N-1}_{j=0}\sigma^x_j$, whose ground state becomes $\ket{\psi(0)}=\ket{+}^{\otimes N}$ with $\sigma^x\ket{\pm}=\pm\ket{\pm}$. We update the initial Hamiltonian after each iteration as $\tilde{H}_i=H_i + \sum_{j=0}^{N-1}h^b_j(\braket{\sigma^z_j})\sigma^z_j$ with $h^b_j(\cdot)$ a function applied to the expectation value $\braket{\sigma^z_j}$ obtained by sampling the lowest energy-valued solutions previously obtained, using a fraction $\alpha\in(0,1]$ of the lowest-energy samples $E_k$ from $E(\alpha)= (1/\lceil \alpha n_\text{shots}\rceil)\sum_{k=1}^{\lceil \alpha n_\text{shots}\rceil} E_k$, with $E_k\le E_{k+1}$~\cite{Barkoutsos2020improving,barron2023provableboundsnoisefreeexpectation,romero2024biasfielddigitizedcounterdiabaticquantum}. Given the iterative learning nature of this method, BF-DCQO is a suitable algorithm to study the interplay between different platforms, as SQC paradigm aims. As an illustrative example, given the scalability and short running times of current quantum annealers, we study the benefits of starting the BF-DCQO on IBM quantum hardware with a fast solution provided by D-Wave. Nevertheless, other approaches might be more suitable depending on the problem nature, desired runtimes, and accessibility to hardware, among other factors. 

Finally, for all cases we apply a minimal local-search (LS) that flips bits and only accepts downhill moves. The process is repeated for a low amount of sweeps, potentially correcting bit-flip errors while searching for neighbouring higher quality solutions. This technique has been widely adopted as post-processing for experimental settings~\cite{simen2025branch,chandarana2025runtime,romero2025protein}. To evaluate the performance of our method, we use the approximation ratio $\text{AR} = E(\alpha=1)/E_0$ as metric, with $E_0$ the exact solution obtained using CPLEX~\cite{cplex}.

\emph{Experiments.}---We consider a $156$-qubit spin-glass model including all the nearest-neighbour (NN) terms provided by the heavy-hexagonal coupling map of \textsc{ibm heron} devices up to three-body terms [\eqlabel{eq:hubo}] with randomly chosen Sidon set couplings $h_i,J_{ij},K_{ijk}\in\{ \pm 8/28, \pm 13/28, \pm 19/28, \pm1 \}$~\cite{sidon1932ein,katzgraber2015seeking}. 
These instances contain $156$ one-, $176$ two- and $244$ three-body terms~\cite{sm}. In~\figlabel{fig:sqc_nn_hubo}, we show the results obtained for an instance, where we compare:%
\begin{itemize}
    \item QA with a large amount of resources: using  \textsc{D-Wave Advantage2\_prototype2.6}~\cite{dwave}, $n_\text{shots}=290000$ and annealing time $t_a=\SI{90}{\micro s}$. We apply $3$ sweeps of LS on all the samples.
    \item Ten BF-DCQO iterations: using \textsc{ibm\_kingston}~\cite{ibm}, $n_\text{shots}=5000$ per iteration. We apply $3$ sweeps of LS on the $2200$ lowest-energy samples after each iteration.
    \item SQC as a warm-started BF-DCQO: first, QA with a low amount of resources, $n_\text{shots}=3000$ and $t_a=\SI{90}{\micro s}$ on \textsc{D-Wave Advantage2\_prototype2.6}. We apply $1$ sweep of LS on the lowest-energy sample. Then, one BF-DCQO iteration on \textsc{ibm\_kingston} using $n_\text{shots}=5000$. We apply $3$ sweeps of LS on the $2200$ lowest-energy samples.
\end{itemize}%

For QA experiments, we used the default settings of \textsc{D-Wave Advantage2\_prototype2.6}, whose connectivity is given by the Zephyr graph~\cite{boothby2021zephyr}. The HUBO-to-QUBO conversion was done with the \textsc{dimod} library, increasing the number of variables to $386$. Posterior embedding on hardware was done with the \textsc{D-Wave Ocean SDK}~\cite{Ocean}, which required $678$ qubits in total. While for standalone QA and BF-DCQO experiments we considered a large amount of shots, in SQC only $8000$ were necessary to find the exact solution (see~\tablabel{tab:performance}). So, for SQC approach it was not only possible to obtain better solutions but also with less resources, obtaining the exact solution with $36\times$ ($6\times$) fewer resources than QA (BF-DCQO), which were unable to find it though. Additionally, the SQC results feature a $1.12\%$ ($2.55\%$) enhancement on the best AR over QA (BF-DCQO) results, despite the AR was $1.7\%$ worse than QA. Finally, none of the post-processed results without considering SQC were able to find the exact solution. The SQC protocol succeeded, showcasing that there are cases that benefit of our proposal when running on composed current quantum hardware.

\emph{Conclusion.}---The rapid advancement of quantum computing promises to tackle complex problems out of the reach of classical computers. However, no platform has emerged as the leading one yet. Superconducting qubits, trapped ions, and neutral-atom systems, among others, each offer distinct advantages and face unique limitations. To combine platforms, leveraging their strengths and offsetting weaknesses, might enable more effective solutions with current quantum platforms. Building on this idea, we present sequential quantum computing (SQC), a novel method that aims to solve problems more efficiently through a selective pool of quantum platforms. Without loss of generality, we address an optimization problem involving quantum annealers and digital quantum hardware. We solve a 156-qubit heavy-hexagonal HUBO problem starting by QA on D-Wave~\cite{dwave}, known to provide fast but not so high-quality outcomes, and use their solutions as an initial guess of an optimization routine run on IBM~\cite{ibm}, which features low connectivity and number of qubits but universality, thus non-stoquastic terms can be implemented to seek higher-quality solutions. Led by our results, SQC is capable of providing faster and better solutions despite the number of quantum resources needed being reduced, showcasing its unique potential to tackle intricate problems with current hardware more efficiently.

\emph{Acknowledgments.}---We thank Pranav Chandarana for valuable discussions and support. We thank Michael Falkenthal and Sebastian Wagner for their help with running the experiments via the PLANQK platform.

\bibliography{reference.bib}

\end{document}


\title{\texorpdfstring{Supplementary Material:\\``Sequential Quantum Computing''}{Supplementary Material: "Sequential Quantum Computing"}}

\author{Sebastián V. Romero$^{\orcidlink{0000-0002-4675-4452}}$}
\affiliation{Kipu Quantum GmbH, Greifswalderstrasse 212, 10405 Berlin, Germany}
\affiliation{Department of Physical Chemistry, University of the Basque Country UPV/EHU, Apartado 644, 48080 Bilbao, Spain}
\author{Alejandro Gomez Cadavid$^{\orcidlink{0000-0003-3271-4684}}$}
\affiliation{Kipu Quantum GmbH, Greifswalderstrasse 212, 10405 Berlin, Germany}
\affiliation{Department of Physical Chemistry, University of the Basque Country UPV/EHU, Apartado 644, 48080 Bilbao, Spain}

\author{Enrique Solano$^{\orcidlink{0000-0002-8602-1181}}$}
\affiliation{Kipu Quantum GmbH, Greifswalderstrasse 212, 10405 Berlin, Germany}

\author{Narendra N. Hegade$^{\orcidlink{0000-0002-9673-2833}}$}
\email{narendrahegade5@gmail.com}
\affiliation{Kipu Quantum GmbH, Greifswalderstrasse 212, 10405 Berlin, Germany}
\date{\today}

\begin{abstract}
    In this Supplementary Material we provide further information about the first-order nested commutator calculation used to approximate the adiabatic gauge potential in our experiments on IBM. Additionally, we explain in detail the initial state preparation for the bias-field digitized counterdiabatic quantum optimization (BF-DCQO) algorithm and how the circuits are decomposed into the native gates of IBM hardware. Finally, based on the graph colouring theorem, we introduce a new method with which to efficiently embed all the two- and three-body terms naturally present in the heavy-hexagonal coupling of IBM devices, method that was used in our experiments.
\end{abstract}

\maketitle\vspace{-5mm}%

\renewcommand{\thetable}{S\arabic{table}}
\renewcommand{\theequation}{S\arabic{equation}}
\renewcommand{\thefigure}{S\arabic{figure}}
\renewcommand{\bibnumfmt}[1]{[S#1]}
\renewcommand{\citenumfont}[1]{S#1}

\section{Circuit implementation of the BF-DCQO algorithm}

\subsection{First-order nested commutator calculation and state initialization}

In the following lines we summarize how we perform the first-order nested commutator calculation, key computation when we apply the bias-field digitized counterdiabatic quantum optimization (BF-DCQO)~\cite{cadavid2024biasfielddigitizedcounterdiabaticquantum,romero2024biasfielddigitizedcounterdiabaticquantum, chandarana2025runtime, iskay} algorithm on digital hardware. It is possible to accelerate adiabatic transitions while suppressing undesired excitations by adding an additional term into the adiabatic evolution $H_\text{ad}$~\cite{demirplak2003adiabatic,berry2009transitionless}, such that the new system evolves under $H_\text{cd}(\lambda)=H_\text{ad}(\lambda) + \dot{\lambda}A_\lambda$, with $A_\lambda$ the adiabatic gauge potential (AGP)~\cite{kolodrubetz2017geometry}. Nevertheless, the exact implementation of the AGP is severely impractical due to its many-body structure and the necessity for the knowledge of the full energy spectrum, where we can approximate it by a nested-commutator series expansion $A_\lambda^{(l} = i\sum_{k=1}^l \alpha_k(t)\mathcal{O}_{2k-1}(t)$, with $\mathcal{O}_0(t) = \partial_\lambda H_\text{ad}$ and $\mathcal{O}_k(t) = [H_\text{ad}, \mathcal{O}_{k-1}(t)]$. In the limit $l\to\infty$, this expansion converges to the exact gauge potential. The coefficients $\alpha_k$ are obtained by minimizing the action $S_l=\trace{G_l^2}$ with $G_l=\partial_\lambda H_\text{ad} - i\big[H_\text{ad},A^{(l}_\lambda\big]$. For simplicity, we consider the first-order ($l=1$) nested-commutator term, thus in our circuits we implement $A^{(1}_\lambda=i\alpha_1(t)[H_\text{ad},\partial_\lambda H_\text{ad}]$ with $\alpha_1=-\trace{\mathcal{O}_1\mathcal{O}^\dagger_1}/\trace{\mathcal{O}_2\mathcal{O}^\dagger_2}$. Higher-order terms demand significantly more computational resources, and their complexity increases with the system size as well as the order of the terms involved. 

Once the first-order nested commutator is computed, the digitized counterdiabatic quantum optimization (DCQO)~\cite{hegade2022digitized} algorithm is run iteratively, sampling and postproccessing with a greedy-pass local search algorithm the final circuits to update the bias fields $h^b_i$ of the initial Hamiltonian $H_i = \sum_{j=1}^N (h_j^x\sigma^x_j + h^b_j\sigma^z_j)$, which modifies its corresponding ground state. The smallest eigenvalue of the single-body operator \( [h_j^x \sigma^x_j - h_j^b \sigma^z_j ]  \)  is given by \( \lambda^{\min}_j = -\sqrt{(h^b_j)^2 + (h^x_j)^2} \), and its associated eigenvector is \( \ket{\phi}_j = R_y(\theta_j) \ket{0}_j \), where \( \theta_j = 2\tan^{-1}[(h^b_j + \lambda^{\min}_j)/h^x_j] \). Therefore, the ground state of $H_i$ can be efficiently prepared using a depth-one layer with $N$ y-axis rotations as $\ket{\psi_i} = \bigotimes_{j=1}^{N} \ket{\phi}_j = \bigotimes_{j=1}^{N} R_y(\theta_j)\ket{0}_j$.

\subsection{Circuit decomposition}

An important aspect of preparing and running quantum circuits on hardware is to transpile the required quantum operations according to the corresponding native gate sets provided by the platform. These typically consist of an universal gate set containing several one-qubit gates and a single two-qubit entangling gate. For \textsc{ibm\_kingston}, its native gate set is composed by
\begin{equation}
    X=\begin{bmatrix}0&1\\1&0\end{bmatrix},\quad\sqrt{X}=\frac{1}{2}\begin{bmatrix}1+i & 1-i\\ 1-i& 1+i\end{bmatrix},\quad R_z(\theta)=e^{-i\theta\sigma^z/2},
\end{equation}
with the entangling gate $\text{CZ}=\diag(1,1,1,-1)$. In addition to them, IBM has recently introduced the fractional gates $R_{zz}(\theta)=e^{-i\theta \sigma^z_0\sigma^z_1/2}$ (with $0<\theta\le\pi/2$) and $R_x(\theta)=e^{-i\theta \sigma^x/2}$~\cite{frac} in their Heron-based processors.

In particular, for the 156-qubit NN HUBO solved in the main text we apply the graph colouring theorem to maximally parallelize both the two- and three-body terms, where we use the technique shown in Ref.~\cite{kim2023evidence} for the two-body terms (three depth-one layers are needed) and extend the same idea for the three-body ones (six depth-one layers required [\figlabel{fig:gc_three}]). 
\begin{figure}[H]
    \centering
    \includegraphics[width=0.44\linewidth]{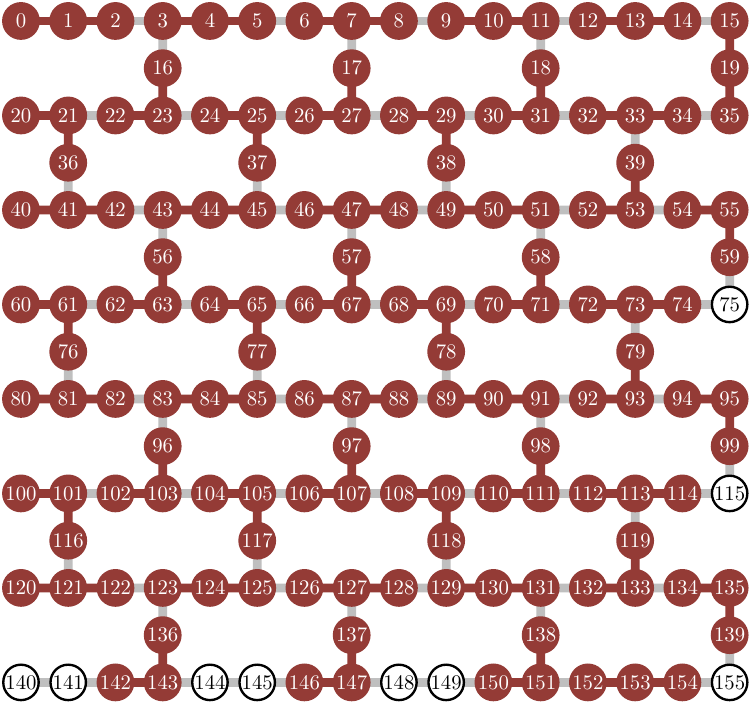}\hspace{5mm}%
    \includegraphics[width=0.44\linewidth]{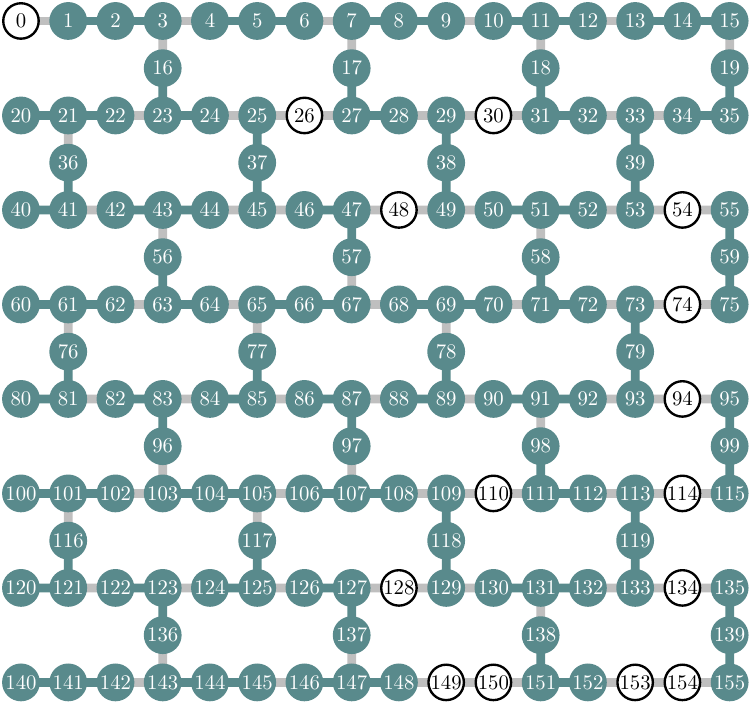} \\ \vspace{1mm}%
    \includegraphics[width=0.44\linewidth]{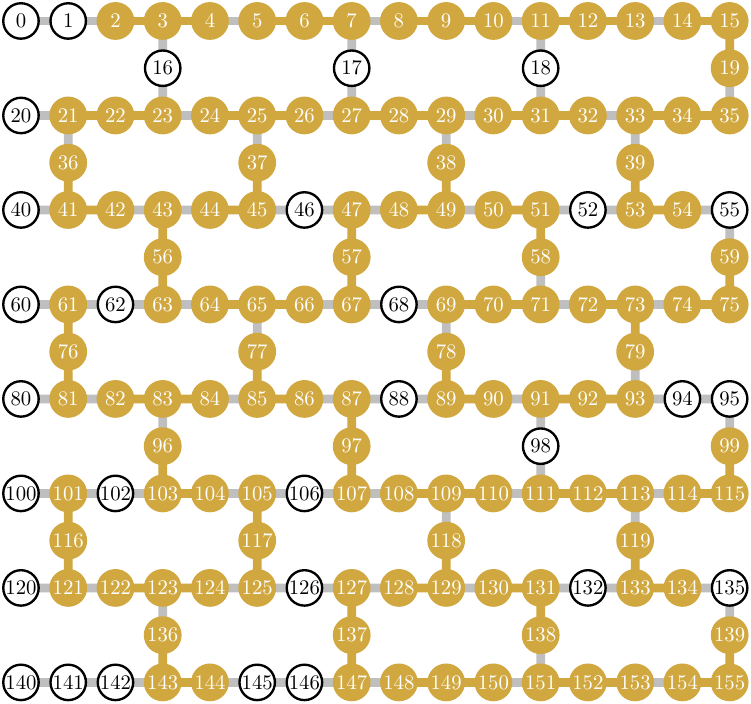}\hspace{5mm}%
    \includegraphics[width=0.44\linewidth]{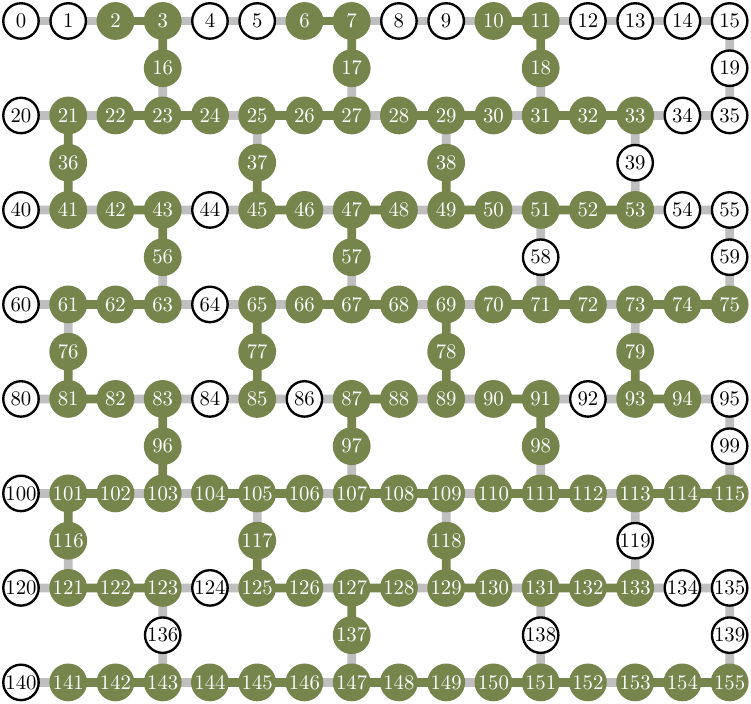} \\ \vspace{1mm} %
    \includegraphics[width=0.44\linewidth]{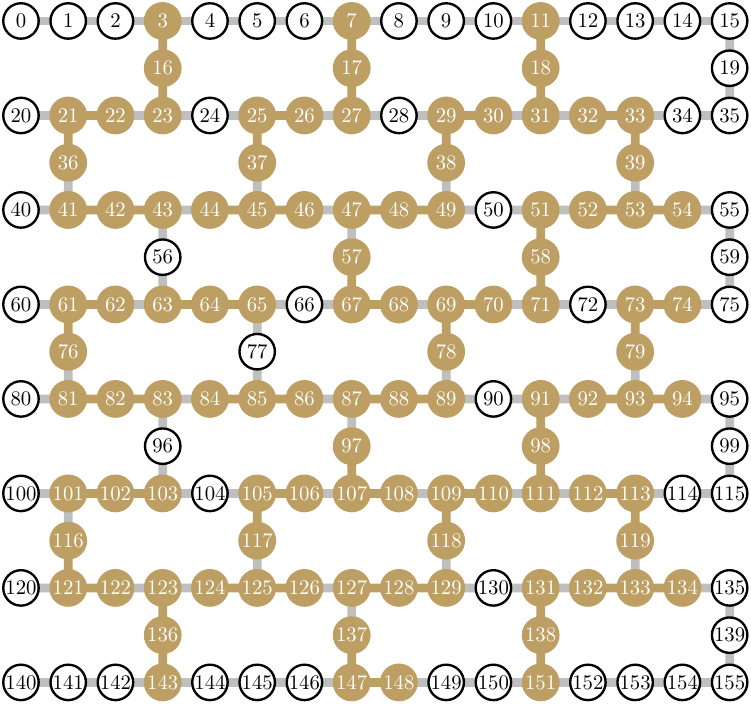}\hspace{5mm}%
    \includegraphics[width=0.44\linewidth]{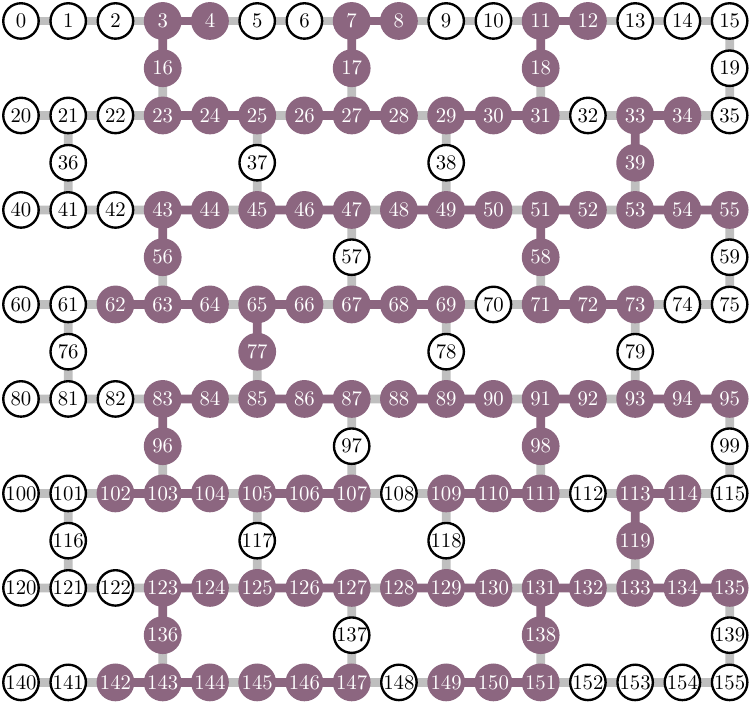}%
    \caption{Six depth-one layers used for solving the 156-qubit NN HUBO in \textsc{ibm\_kingston}. The resulting 244 three-body terms are split into 49, 47, 42, 39, 34, 33 parallelly implemented terms, which are shaded in maroon, teal, mustard, olive, beige and plum colours, respectively. Therefore, each coloured circuit constitutes one out of the six layer blocks to be implemented.}\label{fig:gc_three}%
\end{figure}

\bibliography{reference}